\documentclass[sigconf]{acmart}

\setcopyright{none}
\renewcommand\footnotetextcopyrightpermission[1]{}

\usepackage{fancyhdr}
\fancypagestyle{plain}{%
   \fancyhf{} %
}
\fancypagestyle{firstpagestyle}{%
   \fancyhf{} %
}

\usepackage{enumitem}
\usepackage{balance}
\usepackage{booktabs}
\usepackage[colorinlistoftodos,prependcaption,textsize=footnotesize]{todonotes}
\usepackage{xargs}
\definecolor{OliveGreen}{rgb}{0,0.6,0}
\newcommandx{\answer}[2][1=]{\todo[inline,linecolor=OliveGreen,backgroundcolor=OliveGreen!25,bordercolor=OliveGreen,#1]{#2}}

\usetikzlibrary{calc}
\DeclareRobustCommand{\circled}[1]{%
  {\tikz[baseline=(char.base)]{\node[shape=circle,draw,inner sep=0.6pt] (char) {\fontsize{7pt}{7pt}\selectfont #1};}}\hspace{-2pt}
}

\begin{document}

\title{The Graph Neural Networking Challenge:\\A Worldwide Competition for Education in AI/ML for Networks}

\author{José Suárez-Varela\textsuperscript{1*}, Miquel Ferriol-Galmés\textsuperscript{1}, Albert López\textsuperscript{1}, Paul Almasan\textsuperscript{1}, Guillermo Bernárdez\textsuperscript{1}, David Pujol-Perich\textsuperscript{1}, Krzysztof Rusek\textsuperscript{1 2}, Loïck Bonniot\textsuperscript{3 4}, Christoph Neumann\textsuperscript{3}, François~Schnitzler\textsuperscript{3}, François Taïani\textsuperscript{4}, Martin Happ\textsuperscript{5 6}, Christian Maier\textsuperscript{5}, Jia Lei Du\textsuperscript{5}, Matthias~Herlich\textsuperscript{5}, Peter Dorfinger\textsuperscript{5}, Nick Vincent Hainke\textsuperscript{7}, Stefan Venz\textsuperscript{7}, Johannes Wegener\textsuperscript{7}, Henrike~Wissing\textsuperscript{7}, Bo Wu\textsuperscript{8}, Shihan Xiao\textsuperscript{8}, Pere Barlet-Ros\textsuperscript{1}, Albert Cabellos-Aparicio\textsuperscript{1}}
    \affiliation{ 
      \institution{\textsuperscript{1}Barcelona Neural Networking center, Universitat Politècnica de Catalunya, Spain}
      \institution{\textsuperscript{2}AGH University of Science and Technology, Department of Telecommunications, Poland}
      \institution{\textsuperscript{3}InterDigital, France}
      \institution{\textsuperscript{4}Univ. Rennes, Inria, CNRS, IRISA, France}
      \institution{\textsuperscript{5}Salzburg Research Forschungsgesellschaft mbH, Austria}
      \institution{\textsuperscript{6}IDA Lab, University of Salzburg, Austria}
      \institution{\textsuperscript{7}Fraunhofer HHI, Germany}
       \institution{\textsuperscript{8}Network Technology Lab., Huawei Technologies Co., Ltd., China}
       \institution{\vspace{-0.3cm}\textsuperscript{*}Corresponding author. Email: jsuarezv@ac.upc.edu}
    }

\begin{abstract}
During the last decade, Machine Learning (ML) has increasingly become a hot topic in the field of Computer Networks and is expected to be gradually adopted for a plethora of control, monitoring and management tasks in real-world deployments. This poses the need to count on new generations of students, researchers and practitioners with a solid background in ML applied to networks. During 2020, the International Telecommunication Union (ITU) has organized the ``ITU AI/ML in 5G challenge'', an open global competition that has introduced to a broad audience some of the current main challenges in ML for networks. This large-scale initiative has gathered 23 different challenges proposed by network operators, equipment manufacturers and academia, and has attracted a total of 1300+ participants from 60+ countries. This paper narrates our experience organizing one of the proposed challenges: the ``Graph Neural Networking Challenge 2020''. We describe the problem presented to participants, the tools and resources provided, some organization aspects and participation statistics, an outline of the top-3 awarded solutions, and a summary with some lessons learned during all this journey. As a result, this challenge leaves a curated set of educational resources openly available to anyone interested in the topic.
\end{abstract}



\maketitle

\section{Introduction}\label{sec:intro}

The last decade has witnessed a prolific era of Machine Learning (ML) with a plethora of groundbreaking applications in many fields such as Computer Vision, Natural Language Processing (NLP) or Automated Control (e.g., self-driving cars)~\cite{survey-dl,survey-cars}. This has raised a large interest among the networking community, which is actively investigating on the use of ML techniques to efficiently solve, not only classic networking problems (e.g., Traffic Engineering~\cite{xu2018experience}, Congestion control~\cite{drl-tcp}), but also others related to emerging paradigms (e.g., SDN/NFV~\cite{pei2019optimal}, Internet of Things~\cite{survey-iot}, 5G and beyond~\cite{survey-wireless}). The use of Artificial Intelligence (AI) is thus expected to be ubiquitous in the next-generation Internet, shaping solutions for network control and management, traffic monitoring and analysis, security, or QoS provisioning among many others~\cite{ml-networking,survey-ml-net,itu-imt2020}.

With this prospect, the networking community will arguably experience in the next few years a great demand of professionals with deep knowledge in AI/ML, and special focus on the application of these technologies to computer networks. This requires to prepare new generations of networking students, researchers and practitioners with specific educational programs on AI/ML concepts, while offering the needed motivation and support to help them dive into these subjects. In this vein, the organization of open ML competitions is becoming increasingly popular to raise interest on relevant challenges of a particular area (e.g., Computer Vision, NLP)~\cite{kaggle,driven-data}. This represents a unique opportunity to reach a broad audience via an attractive educational format, where participants are motivated by the incentives of competing with other members and gain recognition from their community (e.g., certificates, public award ceremony, cash prizes). Overall, competitions represent a good alternative, complementary to formal education, that can help attract interest through a fun self-learning experience assisted by the tools and resources provided by the challenge organizers.

As a side contribution, ML competitions may serve to establish reference benchmarks for future research in the field. The availability of public data is crucial to make progress in ML applications, as data is the basic unit to train and evaluate ML models. An example of this is the well-known ImageNet dataset~\cite{imagenet} used in the ILSVRC competition for image classification~\cite{ILSVRC}, which ushered in a series of pioneering breakthroughs in the area of computer vision. In the field of computer networks, it is often difficult to find open datasets for ML applications, as data is typically difficult and costly to obtain, and sharing data with other researchers is often problematic due to privacy issues; hence the lack of public datasets often represents a main barrier to create ML-based solutions for networks. In this context, ML competitions may help establish relevant open datasets for benchmarking existing solutions in the area.

During 2020, the International Telecommunication Union (ITU) has run a competition called the ``ITU AI/ML in 5G challenge''~\cite{itu-challenge}, with the purpose of bringing to light a collection of relevant use cases on ML applied to networking. This competition started with a first round including 23 different challenges that ran in parallel, proposed by network operators, equipment vendors and academia. Then, the winning solutions from these challenges were uniformly evaluated in a final round. By the end of the event, statistics present a total of 1300+ participants from 60+ countries, including a side program with 26 webinars in AI/ML for networks given by experts in the field, and cash prizes for the top teams~\cite{itu-news}.

This paper presents an overview of our experience organizing one of the challenges that have run under this initiative, namely the ``Graph Neural Networking challenge 2020''~\cite{gnnet-challenge}. This challenge presents a fundamental problem related to network performance modeling: predicting the per-path mean delay given a network snapshot (topology + traffic matrix + routing), where only solutions based on neural networks were accepted. This 6-month competition has attracted 124 participants from 24 different countries, including a wide variety of profiles (undergraduate, PhD students, senior researchers and professionals). As a baseline, we provided RouteNet~\cite{routenet} \mbox{-- a} network modeling tool based on Graph Neural \mbox{Networks (GNN)~\cite{scarselli2008graph} --} which is the responsible for the name of the challenge. Overall, the central piece of the competition is a large collection of datasets for network modeling generated with a packet-level network simulator (\mbox{OMNeT++~\cite{omnet}}). We hope these datasets have not only contributed to the development of this competition, but also that they will serve as a reference for future research and education in the field. In short, this challenge leaves the following set of tools and educational resources that can be useful in the future: \mbox{$(i)$ datasets} for network modeling including a wide variety of samples with different topologies, routing configurations and traffic, \mbox{$(ii)$ an} API to easily read and process the datasets, \mbox{$(iii)$ an} open-source implementation of the GNN baseline solution~\cite{routenet} with a quick-start tutorial, and \mbox{$(iv)$ a} webinar presenting the proposed problem and some guidelines to participate in the challenge.

The remainder of this paper is structured as follows: Section~\ref{sec:problem} describes the problem statement of the challenge. Section~\ref{sec:resources} outlines the tools and resources provided to participants. Section~\ref{sec:organization} presents the main organization details and some participation statistics. In Section~\ref{sec:solutions}, the top-3 awarded teams describe briefly their solutions. Section~\ref{sec:lessons-learned} shares some lessons learned during the celebration of the challenge, and finally Section~\ref{sec:Conclusions} concludes the paper.

\section{Problem statement}\label{sec:problem}

Network modeling is essential to build optimization tools for computer networks. A network model is intended to predict the performance (e.g., utilization, delay, jitter, loss) given a network configuration. As a result, these tools can be used to find the optimal configuration (e.g., routing) according to a target policy (e.g., minimize the end-to-end delay). This can have many applications for offline network optimization (e.g., what-if analysis, planning, troubleshooting), but it is also essential to achieve online operation. Particularly, automatic optimization can be achieved by combining a network model with any optimization algorithm (e.g., Local Search~\cite{gay2017expect}, Reinforcement Learning~\cite{xu2018experience}). In this symbiotic combination, the optimization algorithm generates alternative configurations (e.g., routing), while the network model evaluates the resulting performance of each configuration~\cite{routenet}.

A fundamental aspect of optimization tools is that they can only optimize metrics that the network model can predict. For instance, if we want to optimize directly the end-to-end delay in a network (e.g., to meet the customer SLAs), we need a model that can infer the resulting delay with alternative configurations (e.g., routing). In general, if we want to optimize QoS/SLA metrics (e.g., delay, jitter), we need to count on a network model able to predict these metrics for any input configuration. The ultimate ambition of the network modeling community is to achieve an exact digital replica that can reproduce the behavior of real networks, that is why these models are often referred to as Digital Twins (DT)~\cite{dt-ietf}.

Traditional modeling techniques do not meet the requirements to accurately estimate end-to-end performance metrics (e.g., delay or jitter) at limited cost. Nowadays, network models are mainly based on packet-level simulation (e.g., OMNeT++~\cite{omnet}, ns-3~\cite{ns3}) or analytical models (e.g., queuing theory~\cite{queuingModels}). The former are computationally very expensive, while the latter are sufficiently fast but do not produce accurate estimates in real networks with multi-hop routing and multi-queue scheduling policies~\cite{xu2018experience,rusek2020routenet}.

In this context, ML arises as a promising technique to build accurate DTs that can operate in real time. However, existing ML-based solutions do not meet the needs for accurate modeling of real-world networks. A main open challenge lies into the lack of ability of these models to generalize to network scenarios unseen during the training phase (e.g., new topologies, configurations, traffic). Generalization is thus an important property, as it enables to train a DT in a controlled testbed (e.g., in a networking lab) and then be able to commercialize it for direct deployment in any real network, without the need for (re)-training on premises.

\begin{figure}[!t]
\centerline{\includegraphics[width=1.0\columnwidth]{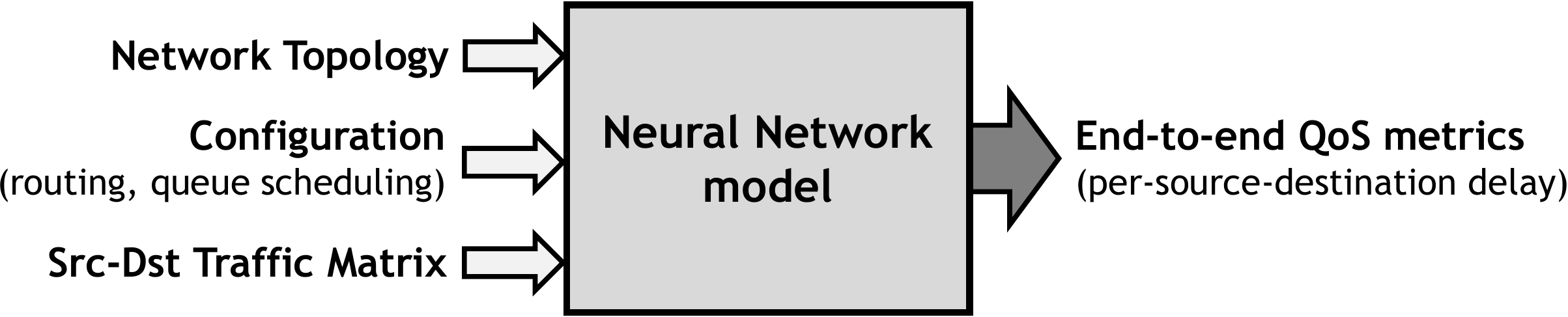}}
\caption{Description of the problem statement.}
\label{fig:problem-statement}
\end{figure}

The goal of the ``Graph Neural Networking challenge 2020'' is to find DT solutions based on neural networks that can accurately estimate end-to-end performance metrics given a network snapshot. Particularly, requested solutions should predict the per-path mean delay given: \mbox{$(i)$ a} network topology, \mbox{$(ii)$ a} network configuration (routing, queue scheduling), and \mbox{$(iii)$ a} source-destination traffic matrix -- as illustrated in Figure~\ref{fig:problem-statement}.

The ambition of this challenge is to test the generalization capabilities of proposed solutions. To this end, we provide a training dataset with a great variety of network scenarios (i.e., topology + configuration + traffic), and then evaluate the accuracy of solutions on another dataset with samples of a different network topology.

\section{Tools and resources}\label{sec:resources}
This section describes the tools and resources that were prepared and publicly released for this challenge. We hope this material can be useful for future education in ML for networks, as well as serving as a reference for research in the area of network modeling. We summarize below the main resources provided:

\begin{itemize}[leftmargin=0.4cm]
\item Datasets for training, evaluation and test (totaling 550k samples).

\item Python API to read and process the datasets.

\item Baseline solution: open-source implementation of RouteNet~\cite{routenet} in TensorFlow v2.1, including a quick-start tutorial.

\item Webinar including a motivation of the proposed problem, some background on Graph Neural Networks, and an overview of their main applications to Computer Networks (presented in the ITU AI for Good Global Summit 2020~\cite{itu-ai4good}).

\item Platform for automatic and real-time evaluation of participants (see more details about the evaluation process in Section~\ref{sec:organization}).

\item Mailing list for questions and answers related to the challenge.

\end{itemize}

All these resources can be accessed via the challenge website~\cite{gnnet-challenge}; except for the evaluation platform, which was closed after the end of the competition. The following subsections provide a more detailed description of the datasets and the baseline solution.

\subsection{Datasets}\label{subsec:datasets}

\begin{figure}[!t]
\centerline{\includegraphics[width=0.9\columnwidth]{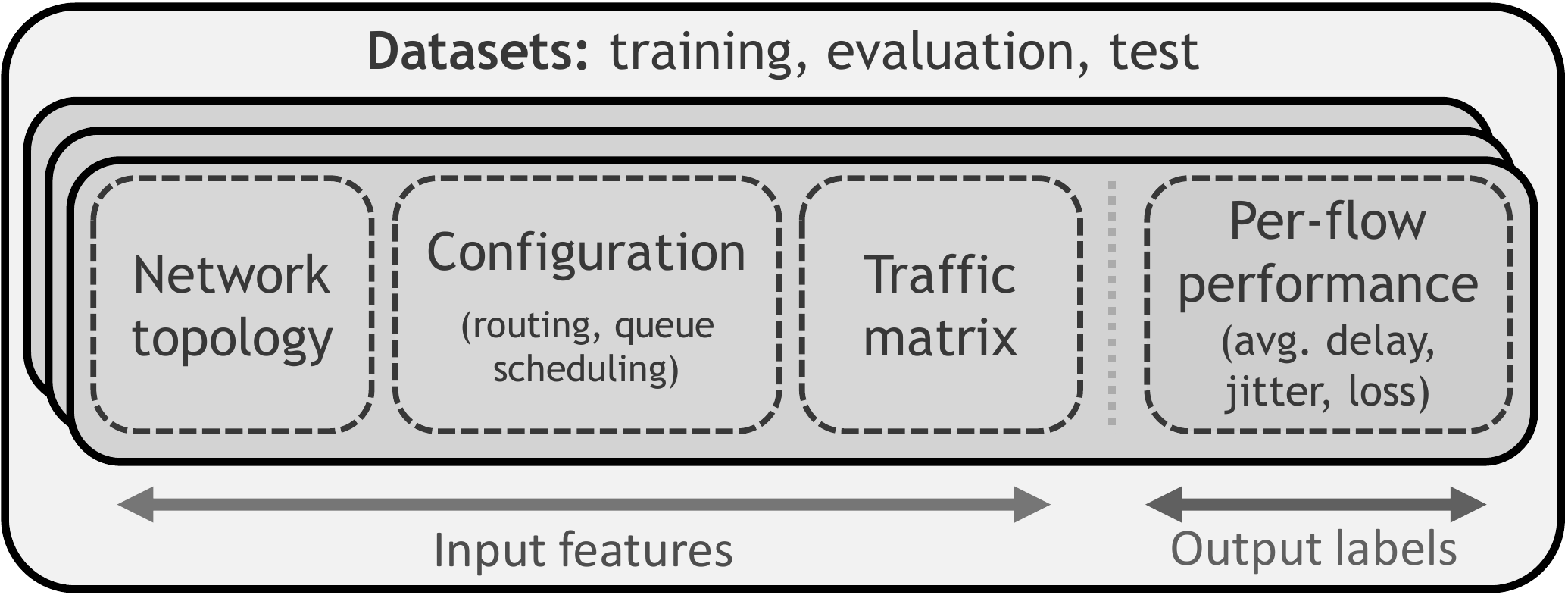}}
\caption{Representation of the dataset.}
\label{fig:dataset}
\vspace{-0.2cm}
\end{figure}

The datasets were produced with a packet-accurate network simulator (OMNet++~\cite{omnet}). Each sample was simulated in a network scenario defined by: $(i)$ a topology, $(ii)$ a routing and queue scheduling configuration, and $(iii)$ a src-dst traffic matrix. Then, it was labeled with some per-flow performance measurements obtained by the simulator (see Fig.~\ref{fig:dataset}). Note that the challenge focuses only on predicting the average per-packet delay at the flow-level glanularity. However, datasets include other relevant per-flow performance metrics that can be useful for future research, such as jitter (i.e., delay variance), some percentiles of the delay distribution (p10, p20, p50, p80, p90), or the amount of packets dropped.

Three different datasets were provided for the competition: \linebreak \mbox{$(i)$ training}, \mbox{$(ii)$ validation}, and \mbox{$(iii)$ test}. As the challenge focuses on the generalization capabilities of proposed solutions, the training dataset contains samples simulated in two real-world network topologies: NSFNet (14 nodes) and Geant2 (24 nodes); while the validation and test datasets were simulated in other topologies not used for training: GBN (17 nodes) for validation, and RedIRIS (19 nodes) for test. All these topologies can be visualized at~\cite{gnnet-dataset}. The training and validation datasets were released at the beginning of the competition, and they included all the per-flow performance measurements (output labels in Fig.~\ref{fig:dataset}). The test dataset did not include any performance measurements, only the input features, and it was released at the end of the competition to evaluate the accuracy of participants' solutions (more details about the evaluation process in Sec.~\ref{sec:organization}). Note that this latter dataset was carefully generated to ensure that input and output features followed a similar distribution to the samples of the validation dataset. For the sake of transparency, we released a new version of the test dataset including the per-flow delay labels after the celebration of the competition.

These datasets combine for each topology hundreds of configurations (i.e., routing, queue scheduling) and traffic matrices with various load levels; totaling 400k, 100k, and 50k samples respectively for training, validation and test. Particularly, routing configurations were generated as variations of the shortest path policy. For the queue scheduling configuration, devices implement different policies: Strict Priority (SP), Weighted Fair Queueing (WFQ), or Deficit Round Robin (DRR). Each of the three datasets are composed of four different subsets --each accounting for 25\% of the samples-- differentiated by the queuing policies supported on devices, which ultimately represent four increasing levels of difficulty. For instance, in the first subset all devices implement a WFQ policy with 3 queues in each output port, and all ports have the same queue weight distribution (10,30,60); while the second subset contains configurations with variable queue weight distributions across different devices. In order to map packets to queues, src-dst flows in the input traffic matrix may have 3 different Type-of-Service labels (ToS=[0,1,2]), which are respectively mapped to the three queues at output ports of devices (e.g., ToS=0 is mapped to the first queue). 

Lastly, traffic matrices are carefully generated to cover a wide range of traffic intensities. To this end, for each sample the simulator randomly selects a reference maximum traffic intensity level ($TI_{max}$), from low to high load (400-2000 bits per time unit in the simulator). Then, the traffic rate of src-dst flows [$TM(src,dst)$] is sampled using a uniform distribution~($\mathcal{U}$):
\vspace{-0.1cm}
$$TM(src,dst) = \mathcal{U}(0.1,1)*TI_{max}\vspace{-0.1cm}$$

During simulation, inter-packet arrival times are modeled with an exponential distribution, and packet size follows a bimodal distribution; both adapted to fit the rate of each src-dst flow. We refer the reader to~\cite{gnnet-dataset} for more details about the datasets.

\subsection{Baseline}\label{subsec:baseline}

Graph Neural Networks (GNN)~\cite{scarselli2008graph} have recently shown a strong potential to generalize over graph-structured information~\cite{battaglia2018relational}. This makes these models especially interesting for networking applications, as much network-related information is fundamentally represented as graphs (e.g., topology, routing). Particularly, global control and management problems (e.g., routing optimization) are often formulated as $\mathcal{NP}$-hard combinatorial optimization problems~\cite{vesselinova2020learning} that involve a set of network elements with complex relationships between them (i.e., graphs). For instance, to account for QoS metrics (e.g., flow-level delay), we need to model the inter-dependencies between flows in the network, and consider the state of the links and devices that these flows traverse --which in turn depends on the routing configuration. All these dependencies are non-trivial to model, as they are circular --i.e., mutually recursive.

Nowadays, GNN is a hot topic in the ML field and, as such, we are witnessing significant efforts to leverage its potential in different fields where data is fundamentally represented as graphs (e.g., chemistry, physics, social networks)~\cite{zhou2018graph}. In the networking field, early works~\cite{routenet,geyer2018learning,almasan2019deep} have already demonstrated the potential of GNNs to generalize across different network scenarios. This can be interesting from a commercialization standpoint, as it enables to train ML-based solutions in controlled testbeds (e.g., in a networking lab), and then be able to deploy them in real networks in production, without the need for on-site training. We refer the reader to ~\cite{scarselli2008graph, battaglia2018relational, gilmer2017neural} for more generic background on GNN.

\begin{figure}[!t]
\centerline{\includegraphics[width=0.9\columnwidth]{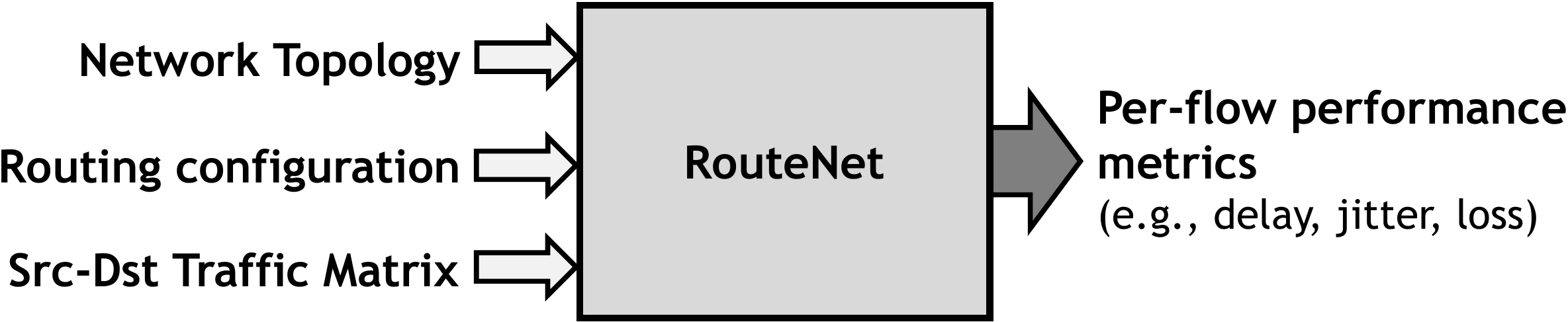}}
\caption{Black-box representation of RouteNet.}
\label{fig:routenet}
\vspace{-0.6cm}
\end{figure}

As a baseline, we presented to participants RouteNet~\cite{routenet}, a GNN model intended to estimate flow performance metrics in networks (e.g., delay, jitter, loss). Thanks to the use of GNN, RouteNet revealed ability to make accurate predictions in networks unseen during the training phase, including other network topologies, routing configurations, and traffic matrices (Fig.~\ref{fig:routenet}). We refer the reader to the original paper~\cite{routenet} for more details about this model.
              
In the Graph Neural Networking Challenge, we extended the problem to modeling network performance (per-flow delay) in scenarios with various queue scheduling policies across network devices (Fig.~\ref{fig:problem-statement}). In this vein, RouteNet does not support multi-queue scheduling policies as input features; moreover it is not designed to map traffic flows to particular queues according to their Type-of-Service labels, which would be helpful to estimate accurately the per-flow delay in the network scenarios of the datasets (Sec.~\ref{subsec:datasets}). 

We provide an open source implementation of RouteNet (TensorFlow v2.1)~\cite{gnnet-baseline}, along with a tutorial on how to execute it and modify key features of the model. Participants were encouraged to use this model as a starting point to solve the challenge, or alternatively create their own neural network-based solutions from scratch.

\section{Organization and participation statistics}\label{sec:organization}

At the beginning of the challenge, participants were given the training and validation datasets~(Sec.~\ref{subsec:datasets}), an open-source implementation of the baseline model (Sec.~\ref{subsec:baseline}), and a webinar presenting the problem. This latter activity was intended to engage participants and guide them on how to start working for the competition. For this purpose, we motivated the relevance of the problem and introduced the main existing solutions; also we gave some tips to be successful in the challenge. Moreover, during the competition we actively communicated with participants via the mailing list.

Approximately 4 months after the challenge kickoff, we released the test dataset, which marked the start of the evaluation phase. This dataset contained samples with similar distributions to those of the validation dataset. Participants were asked to label it with their neural network models and return the results in a standard CSV format. This enabled to make the evaluation in real time, as our evaluation platform automatically computed the resulting scores over CSV files. Particularly, the score was defined as the Mean Absolute Percentage Error (MAPE) over all the source-destination delay predictions produced by solutions (Eq.~\ref{eq:mape}). 
\vspace{-0.2cm}

\begin{equation}
MAPE = \frac{100\%}{N} \sum_{i=1}^{N}{\displaystyle\left\lvert \frac{\hat{y}_i-y_i}{y_i}\right\rvert}
\label{eq:mape}
\end{equation}

Where $\hat{y}_i$ are the delay predictions produced by the model, $y_i$ represents the actual delay labels obtained by the simulator, and \textit{N} is the number of src-dst flows along the 50k samples of the test dataset, totaling 17,100,000 src-dst delay labels. The evaluation phase lasted 15 days and, as solutions were submitted, teams were ranked automatically by the evaluation platform according to their MAPE metric (less is better). They could see the evolution of the ranking in real time, which helped create a competitive atmosphere and get the participants more involved.

At the end of the evaluation phase, a provisional ranking with the scores of all the teams was publicly posted; then top-5 solutions were analyzed in detail to check that they complied with all the rules of the competition. Among the most essential aspects, we checked that these solutions were fundamentally based on a neural network model, and did not use for instance any simulation tool as the one used to generate the datasets. Also, as the challenge was focused on the generalization capabilities of solutions, we imposed that solutions could only be trained with samples from the training dataset we provided, thus ensuring equal conditions for all the teams with regard to data. This led us to reproduce the training and evaluation of top-5 solutions to check that they really met this rule.

After this validation, top-3 participants were officially announced and received certificates of recognition. Also, they advanced to the global round of the ITU AI/ML in 5G challenge, where 33 solutions from 23 different challenges were uniformly evaluated by an expert panel during a final public conference. The global winners received certificates of recognition (gold, silver and bronze medals, and three runner-up awards) and the corresponding cash prizes~\cite{itu-news}.

\begin{table}[!b]
\vspace{-0.2cm}
\caption{List of countries represented in alphabetical order.} \label{table:countries}
\begin{tabular}{p{0.9\columnwidth}}
\toprule
\multicolumn{1}{c}{\textbf{Countries}}\\
\midrule
Austria, Belgium, China, Colombia, Egypt, Finland, France, Germany, Greece, India, Israel, Italy, Jamaica, Japan,\\Luxembourg, Malaysia, Portugal, Russia, Saudi Arabia,\\South Africa, Spain, Tunisia, United Kingdom, United States\\
\bottomrule
\end{tabular}
\end{table}

The whole challenge lasted 6 months (7 months including the global round), and attracted a total of 124 participants from 24 countries, including a wide variety of profiles: undergraduate, PhD students, senior researchers, professionals; both from the academia and industry. Table~\ref{table:countries} summarizes the countries represented across all the teams. Also, by manual inspection of the affiliations they provided, we identified that $\approx$59\% of participants came from the academia (universities and research centers/institutes), while $\approx$24\% were researchers and professionals from the industry. Note that a small portion of them were not considered in any of these groups; likewise some had dual affiliations that counted for both profiles.

\section{Overview of top-3 solutions}\label{sec:solutions}
This section presents the top-3 winning solutions of the challenge, which were promoted to the global round of the ITU AI/ML in 5G challenge. Table~\ref{table:ranking} summarizes the MAPE scores they obtained. As a reference, the baseline GNN model (Section~\ref{subsec:baseline}) obtains $\approx$269\% of MAPE over the datasets of the challenge (Section~\ref{subsec:datasets}), which reveals a significant improvement of these solutions over the baseline, mainly due to their capability to accurately model the impact of various multi-queue scheduling policies on network devices.

\begin{table}[!t]
\caption{Ranking with top-3 teams.} \label{table:ranking}
\vspace{-0.1cm}
\begin{tabular}{l l l}
\toprule
\textbf{Rank} & \textbf{Team name} & \textbf{MAPE}\\
\midrule
\#1 & Steredeg & 1.53\\
\#2 & Salzburg Research & 1.95\\
\#3 & Gradient Ascent & 5.42\\
\bottomrule
\end{tabular}
\vspace{-0.3cm}
\end{table}

\subsection{First: Steredeg (InterDigital, Univ. Rennes)}

This team observed that node states cannot be modeled explicitly with RouteNet. Their main assumption is that adding embeddings (learned representations) for nodes would allow to store relevant information about node queues (e.g., available capacity, drop rates).
Adding this new embedding is challenging in the RouteNet baseline, so they re-implemented an algorithm from scratch using PyTorch and its GNN module ``PyTorch-Geometric''.
Dependencies between paths', links' and nodes' embeddings were defined using three bipartite graphs:
$(i)$ $G_{p,l}$ connects each path to the links they traverse
--it is essentially the bipartite graph used in RouteNet; \mbox{$(ii)$ $G_{p,n}$} connects each path to the nodes it traverses;
and \mbox{$(iii)$ $G_{l,n}$} connects each (unidirectional) link to its \emph{source} and \emph{destination} nodes.

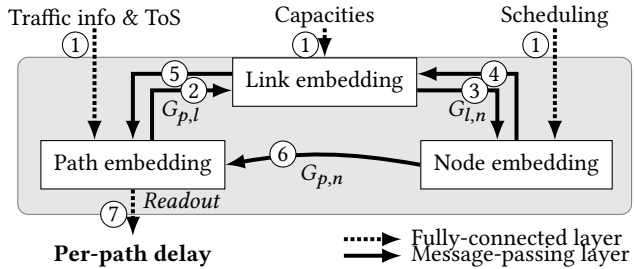
\begin{figure}
  \centering
  \begin{tikzpicture}[x=2.55cm,y=1.1cm]
  \tikzstyle{embedding}=[draw,black,rectangle,inner sep=5pt,fill=white]
  \tikzstyle{update}=[draw,black,->,line width=1.5pt,>=latex]
  \tikzstyle{feature}=[draw,black,->,line width=1.5pt,>=latex,densely dotted]
  \tikzstyle{circled}=[draw,circle,inner sep=1pt,fill=white]
  \definecolor{g}{gray}{.9}

  \node (pathf) at (-.2, 1.65) [anchor=south, inner sep=1pt] {Traffic info \& ToS};
  \node (linkf) at (1, 1.65) [anchor=south, inner sep=0pt] {Capacities};
  \node (nodef) at (2.2, 1.65) [anchor=south, inner sep=0pt] {Scheduling};

  \draw [fill=g,draw=gray,rounded corners] (-.6, 1.3) rectangle (2.6, -.6);

  \node [embedding] (path) at (0, 0) {Path embedding};
  \node [embedding] (link) at (1, 1) {Link embedding};
  \node [embedding] (node) at (2, 0) {Node embedding};

  \draw [feature] (pathf.south) -- (path.north -| pathf.south);
  \draw [feature] (linkf.south) -- (link.north -| linkf.south);
  \draw [feature] (nodef.south) -- (node.north -| nodef.south);

  \draw [update] (path.north) ++(.1,0) |- node [anchor=north west] {$G_{p,l}$} ($(link.west)+(0,-.1)$);
  \draw [update] (link.east) ++(0,-.1) -| node [anchor=north east] {$G_{l,n}$} ($(node.north)+(-.1,0)$);
  \draw [update] (node.north) |- ($(link.east)+(0,.1)$);
  \draw [update] (link.west) ++(0,.1) -| (path.north);
  \draw [update,bend right=10] (node.west) to node [anchor=north] {$G_{p,n}$} (path.east);

  \node (readout) at (0, -1.1) {\textbf{Per-path delay}};
  \draw [feature] (path.south) -- node [anchor=west,yshift=5pt] {\emph{Readout}} (readout.north);

  \draw [feature] (1.1, -.9) -- ++(.3, 0) node [anchor=west] {Fully-connected layer};
  \draw [update] (1.1, -1.1) -- ++(.3, 0) node [anchor=west] {Message-passing layer};

  \draw (pathf.south)+(-.1,-.2) node [circled] {1};
  \draw (linkf.south)+(-.1,-.2) node [circled] {1};
  \draw (nodef.south)+(-.1,-.2) node [circled] {1};
  \draw (link.west)+(-.2,-.1) node [circled] {2};
  \draw (link.east)+(.3,-.1) node [circled] {3};
  \draw (link.east)+(.4,.1) node [circled] {4};
  \draw (link.west)+(-.3,.1) node [circled] {5};
  \draw (path.east)+(.3,.13) node [circled] {6};
  \draw (path.south)+(-.1,-.3) node [circled] {7};
\end{tikzpicture}
  \vspace{-0.6cm}
    \caption{%
    Overview of Steredeg's proposed algorithm.
  }\label{fig:steredeg}
  \vspace{-0.3cm}
\end{figure}

Figure~\ref{fig:steredeg} provides an overview of the interactions between embeddings and the proposed dependency graphs.
Embeddings are first initialized with input features (step \circled{1}).
Using message passing layers over the bipartite graphs $G_{p,l}$ and $G_{l,n}$, the embeddings are then updated in steps \circled{2} to \circled{5} (repeated $T = 3$ times to allow embeddings to converge to a fixed point).
The final embeddings are computed using $G_{p,n}$ \circled{6} and readout layers output the predicted delay for each path \circled{7}.
Numerical input features are standardized while categorical features are mapped to embeddings.
Other improvements include using the Adam optimizer with a cyclic learning rate scheduler (to improve models' convergence) and training over the \emph{logarithm} of the expected delays with the MAPE loss.
The best hyperparameters allowed to reach a score of 1.66\% (5 days of training on a Nvidia Tesla M60).
The winning score of 1.53\% was obtained by averaging the predictions of 4 models trained with various hyperparameters.

\subsection{Second: Salzburg Research (SRFG, Univ. Salzburg)}
This solution is based on the RouteNet model provided for the challenge. Without any modifications, the original model performs poorly as it was not developed for networks with heterogeneous scheduling policies.

The overall structure of RouteNet was kept the same. Two neural networks exchange information with each other in a loop to update path and link state information. These two neural networks are gated recurrent neural networks in RouteNet. In our solution, the path update neural network was extended to a stacked gated recurrent network. The link update neural network was replaced by a feed forward neural network with 2 hidden layers.

Additionally, important variables such as queue scheduling policies were added to the model and all variables were scaled onto the interval $[0,1]$. Note that the scheduling policy is a node property. However, RouteNet only considers path and link information. Hence, we mapped this queue scheduling policy onto links by assigning to each link the scheduling policy of its source node. We set the dimension for path and link states to 64 elements each. For the path states, 11 elements are initialized with data, then we apply zero-padding to the vectors. Similarly, link states are initialized with 7 variables taken from the data set.
Additionally, the loss function for training the model was changed from the Mean Squared Error originally used by RouteNet to the Mean Absolute Percentage Error (MAPE), which is the target score of the challenge. 

The final output from the path update function is used for predicting average per-path delays with a feed forward neural network consisting of two hidden layers in RouteNet. An additional layer was added to this readout neural network without  activation function. This layer additionally receives the final path state information as an input. This can be seen as some kind of residual neural network.

The training was done for 1.2 million training steps on a single Geforce RTX 2080 Ti, which took $\approx$48 hours. With this model, a MAPE of 1.95\% was achieved for the test dataset simulated in the RedIRIS topology. The code of this solution can be found on GitHub\footnote{\url{https://github.com/ITU-AI-ML-in-5G-Challenge/PS-014.2_GNN_Challenge_SalzburgResearch}}. An improved version of this model~\cite{happ2021} achieves an error of 0.89\% for the same test data set.

\subsection{Third: Gradient Ascent (Fraunhofer HHI)}

Gradient Ascent's solution is, in essence, a heavily scaled up version of RouteNet. Modifications to the original model are as follows:

\begin{enumerate}[leftmargin=*]
\item The number of input features of both links and paths is increased, in order to accommodate the heterogeneity of links and paths. Compared to the original RouteNet, the set of link features is extended by the link capacity and the queue scheduling configuration, and the queue weight profiles at the ingress port of the link. The set of path features is extended by the ToS label, the avg. packet size, and the avg. packet rate. Where necessary, these features were mapped to a floating point variable.

\item The sizes of the path and link hidden state vectors (embeddings) used in the message passing phase of the GNN are increased by factors of eight and four, respectively.

\item The dimension of the two hidden layers of the fully-connected readout Neural Network is extended. Particularly, the number of nodes in the hidden layers was increased by a factor 64.
\end{enumerate}

Since the validation loss clearly converges after $\approx$1 million training steps with no sign of overtraining,
no dropout layers are being used in the readout Neural Network
(contrary to the original RouteNet) and the L2 regularization is relaxed compared to the original.
Finally, the number of message passing iterations can be lowered by a factor of two without any loss in accuracy, while reducing computation time. Training of the model (1.2 million training steps) takes about 15 hours on an Nvidia Tesla V100.

\section{Lessons learned}\label{sec:lessons-learned}

This section collects some lessons learned before and during the organization of the challenge.

The use of an objective evaluation metric --MAPE in this case-- was positively perceived by participants. First, it enables to achieve a transparent, fair, and uniform way to evaluate solutions. Second, it is an automatic and scalable mechanism, which is essential to avoid potential problems during the evaluation phase. In this context, it is also important to be prepared to resize the computing resources needed, as it is often difficult to foresee the level of participation before the challenge starts. For instance, this can be achieved by relying on cloud computing resources with flexible plans.

One aspect that can be revisited for future competitions is the definition of the rules and how to verify that all solutions comply with them. In this challenge, the evaluation process was divided in two phases; during the first phase solutions were evaluated automatically based on the MAPE score and, in the second phase, top-5 solutions were shortlisted and manually reviewed. We did this to verify to main aspects:  $(i)$~that solutions were essentially based on neural networks and did not include any network simulation component, and $(ii)$~that they were exclusively trained on data from the training dataset. This permitted to ensure that at least \mbox{top-5} solutions complied with all the rules. However, it was not a trivial task, as it required to reproduce the training and evaluation of these solutions. In this vein, alternative strategies could be devised to check this type of rules in a more efficient way. For instance, providing access to sandboxes where participants can autonomously run their solutions in controlled and homogeneous conditions.

One essential aspect for the successful development of the challenge was to actively engage participants. The initial webinar and the use of a Q\&A mailing list were essential to attract participants and keep interest along the challenge. Also, it is essential to provide tools and resources that can facilitate a quick start. For instance, we provided an API for the datasets, an implementation of the baseline model, and a tutorial with some guidelines on how to modify fundamental features of this model. This kind of resources help achieve promising results at an early stage of the competition; thus encouraging participants to keep working until the end.

During the evaluation phase, the publication of a real-time ranking allowed participants to compare their scores; this eventually encouraged them to compete actively. Particularly, we observed a clear last-mile effort at the end of the competition. In this vein, some actions could be made to achieve a more constant dedication along the challenge.

Another interesting aspect was to observe that particularly top-3 teams achieved very close scores by devising substantially different and complementary approaches (e.g., adding new elements to the neural network architecture, adding new variables to the model, hyperparameter optimization). In this context, it would be interesting to broaden the spectrum of accepted solutions in the challenge (e.g., any ML-based technique beyond neural networks) to achieve more diversity in proposed solutions and, at the end of the challenge, prepare a joint activity where teams can collaborate to create a final solution including the main benefits of the best proposals.

Last but not least, in this challenge we used data from simulation, using statistical traffic models, which provides a simplified representation of real-world networks. In this vein, we would like to encourage the community to move towards using real-world data. For example, using controlled network testbeds and injecting real traffic traces under scenarios with a broad variety of configurations (e.g., routing, queue scheduling).

\vspace{-0.1cm}
\section{Conclusions}\label{sec:Conclusions}
This paper has reported our experience throughout the celebration of the Graph Neural Networking Challenge 2020, a worldwide competition on ML for networking that has been organized under a broader initiative: the ITU AI/ML in 5G challenge. We hope our experience can be helpful for the organization of future competitions in the field of computer networks. From a research standpoint, this kind of competitions can serve to accelerate progress in a particular knowledge field, as it helps to introduce the topic to new researchers and establish standard benchmarks. From an education perspective, it is an effective initiative to attract and train new generations of students and researchers in the subject.

\vspace{-0.1cm}
\begin{acks}
Thanks to the International Telecommunication Union (ITU) for organizing the ITU AI/ML in 5G challenge and giving us the opportunity to be part of this initiative; especially thanks to Dr. Thomas Basikolo, Mr. Vishnu Ram and Dr. Reinhard Scholl, who did a great job to coordinate this large-scale event. This work has received funding from the European Union’s H2020 research and innovation programme within the framework of the NGI-POINTER Project funded under grant agreement No. 871528. This paper reflects only the authors' view; the European Commission is not responsible for any use that may be made of the information it contains. This work was also supported by the Spanish MINECO under contract TEC2017-90034-C2-1-R (ALLIANCE), the Catalan Institution for Research and Advanced Studies (ICREA), and by FI-AGAUR  grant  by  the  Catalan  Government. Salzburg Research is grateful for the support by the WISS 2025 (Science and Innovation Strategy Salzburg 2025) project ''IDALab Salzburg'' (20204-WISS/225/197-2019 and 20102-F1901166-KZP) and the 5G-AI-MLab by the Federal Ministry for Climate Action, Environment, Energy, Mobility, Innovation and Technology (BMK) and the Austrian state Salzburg.

\end{acks}

{ \balance
{
	\bibliographystyle{ACM-Reference-Format}
	\bibliography{references}
}
}

\end{document}